\documentclass[a4paper]{mem}
\usepackage{natbib}
\usepackage{graphicx}
\usepackage[a4paper]{hyperref}
\idline{5}{148}
\def\src {RX\,J0806.3+1527}
\def\srcm {RX\,J1914.4+2456}

\newcommand{\AXAF}{{\em Chandra}}
\newcommand{\R}{{\em ROSAT}}

\newcommand{\xmm}{{\em XMM--Newton}}

\begin{document}
   \title{Taking the pulse of the shortest orbital period binary system \src
\thanks{Based on observations collected at the European Southern Observatory, 
Chile (268.D-5737, 070.D-652 and 072.D-0717)}
}

\author{G.L. Israel\inst{1}, S. Covino\inst{2}, S. Dall'Osso\inst{1}, D. Fugazza\inst{2}, C.W. Mouche\inst{3}, L. Stella\inst{1}, S. Campana\inst{2}, V. Mangano\inst{1,4}, G. Marconi\inst{5}, S. Bagnulo\inst{5} \and U. Munari\inst{6}
%\thanks{this is a place for placing a footnote in the author field }
}

   \offprints{GianLuca Israel}
\mail{gianluca@mporzio.astro.it }		

\institute{INAF - Osservatorio Astronomico di Roma, Via Frascati 33,
Monteporzio Catone, Italy \and
%Affiliated to I.C.R.A.\and
INAF - Osservatorio Astronomico di Brera, Via Bianchi 46, Merate, Italy \and
Lawrence Livermore National Laboratory, 7000 East Avenue, Livermore, USA \and
IASF - CNR  Sezione di Palermo, Via Ugo La Malfa 153, 90146 Palermo, Italy \and
European Southern Observatory, Casilla 19001, Santiago, Chile \and 
INAF - Osservatorio Astronomico di Padova, I-36012 Asiago, Italy}

   \abstract{\src\ is thought to be a 321s orbital period (the
shortest known) double white dwarf binary system. According to the
double degenerate binary (DDB) scenario this source is expected to be
one of the strongest Gravitational Wave (GW) emitter candidates.  In
the last years \src\ has been studied in great details, through
multiwavelength observational campaigns and from the point of view of
data analysis result interpretations. We present here the timing
results obtained thanks to a 3.5-year long optical monitoring campaign
carried out by the Very Large Telescope (VLT) and the Telescopio
Nazionale Galileo (TNG) which allowed us to detect and study the
orbital period derivative (spin--up at a rate of $\sim$10$^{-3}$s
yr$^{-1}$) of the 321s modulation, to detect the linear polarisation
(at a level of about 2\%), and to study the broad band energy
spectrum. The VLT/TNG observational strategy we used allowed us to
rely upon a P-\.P coherent solution which we finally extended backward
to the 1994 ROSAT observation of \src.

   }
   \authorrunning{G.L. Israel et al.}
   \titlerunning{Taking the pulse of the fastest binary system}
   \maketitle
%
%________________________________________________________________

\section{Introduction}

%----------------------------------------------------------- S_vib   
   \begin{figure*}
   \centering
   \resizebox{\hsize}{!}{\includegraphics{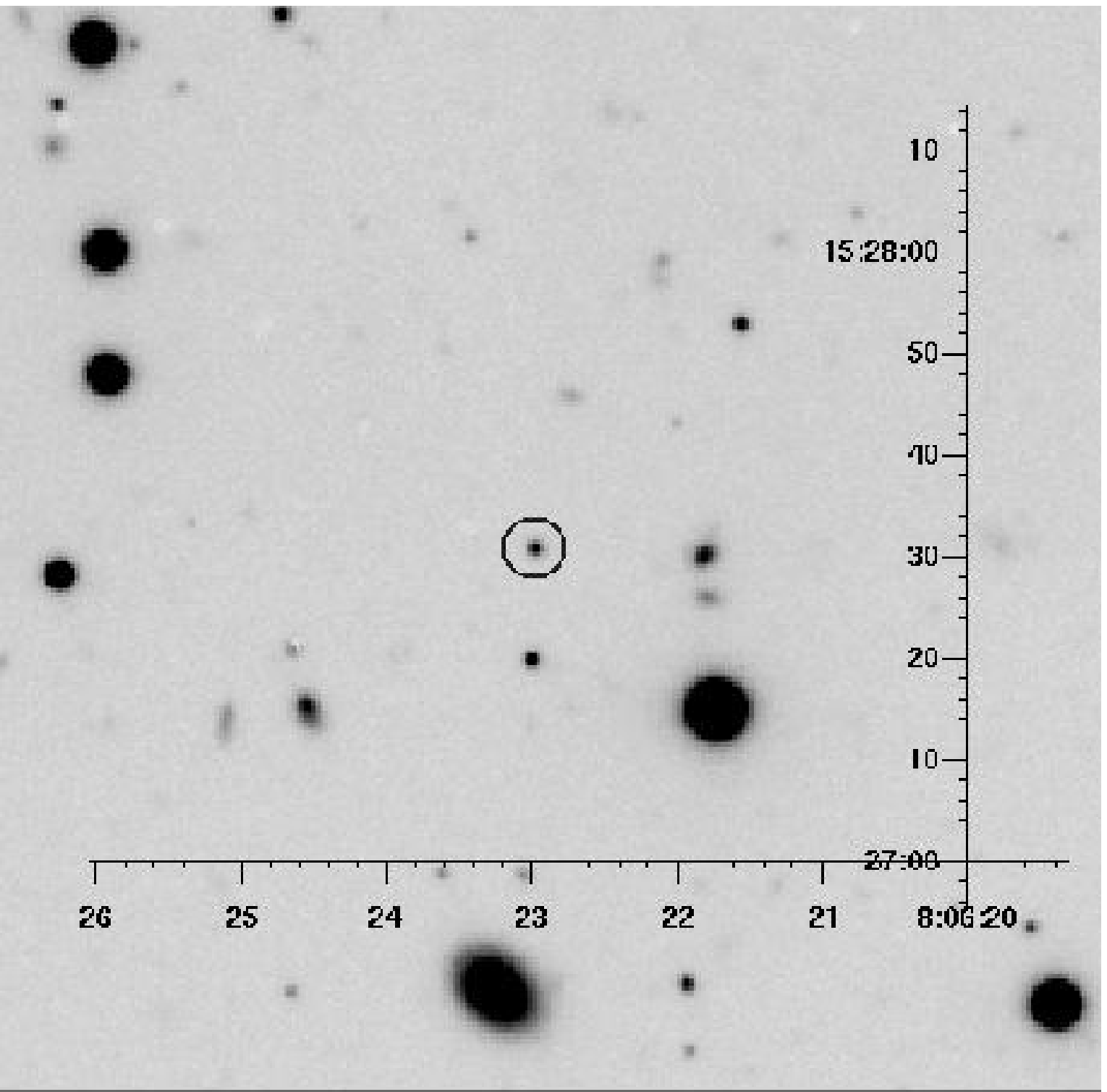}
\includegraphics{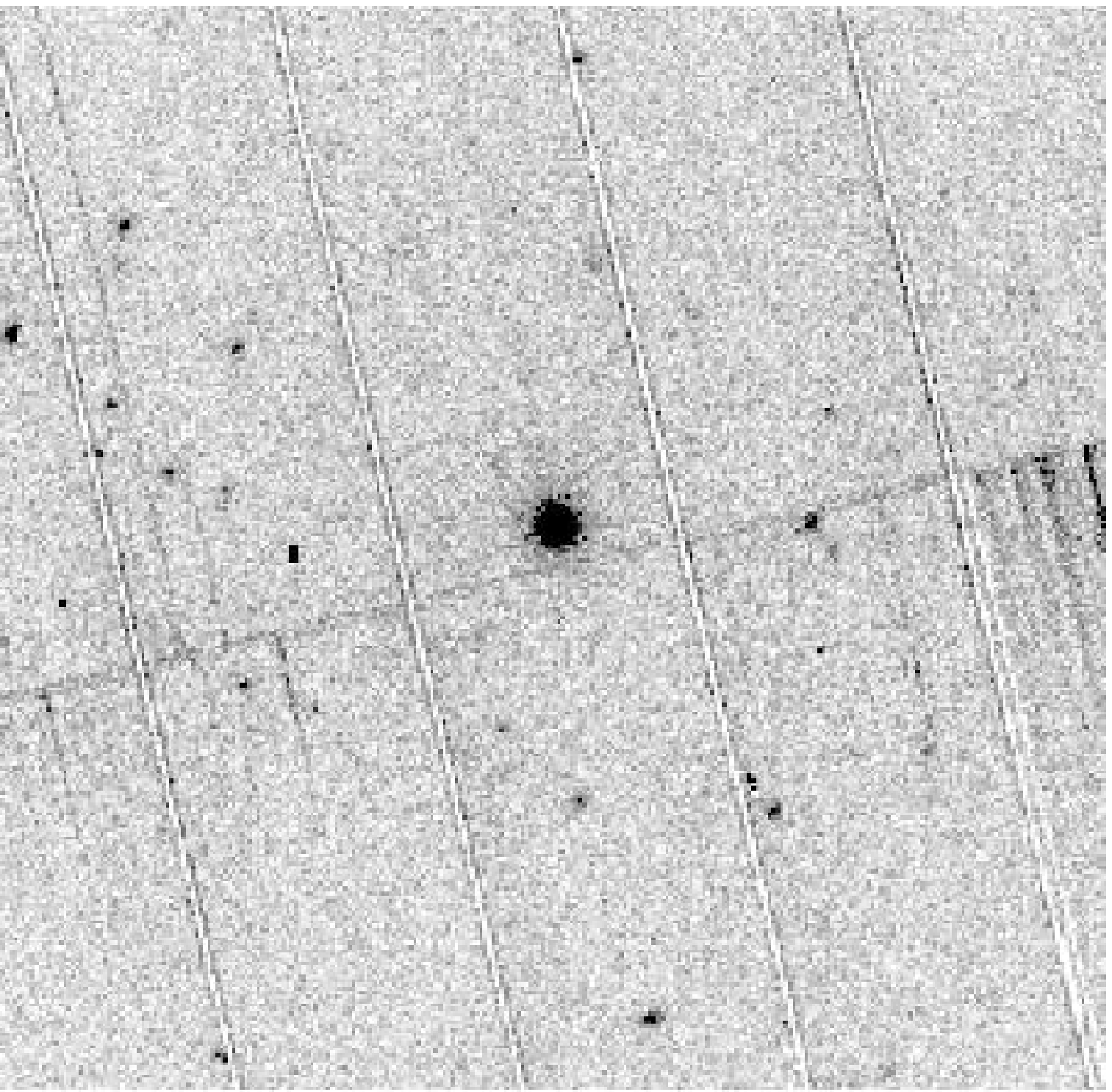}}
    \caption{Left panel: March 2003 VLT FORS1 R-band image of the sky
region including the position of the optical counterpart to \src\
(marked with the circle; effective exposure time of 40 minutes). Right
panel: 0.2-10.0\,keV EPIC-pn \xmm\ image of \src\ (brightest object;
exposure time of $\sim$26000\,s). }
        \label{images}
    \end{figure*}
%
%______________________________________________________________
%%

\src\ was discovered in 1990 with the \R\ satellite during the All-Sky
Survey \citep[RASS;][]{beu99}. However, it was only in 1999 that a
periodic signal at 321\,s was detected in its soft X-ray flux with the
\R\ HRI \citep[][hereafter I99]{io99} \defcitealias{io99}{I99}. The detection 
of X-ray
pulsations was also reported independently by \citet{bur01}.
Subsequent deeper optical studies allowed to unambiguously identify
the optical counterpart of \src, a blue $V=21.1$ ($B=20.7$) star
\citep[][hereafter I02B]{io02a,io02b}\defcitealias{io02b}{I02B}. $B$,
$V$ and $R$ time-resolved photometry revealed the presence of a $\sim
15$\% modulation at the $\sim 321$\,s X-ray period \citep{io02b,ram02a}.
%                                                Two column figure
%----------------------------------------------------------- S_vib   
   \begin{figure*}
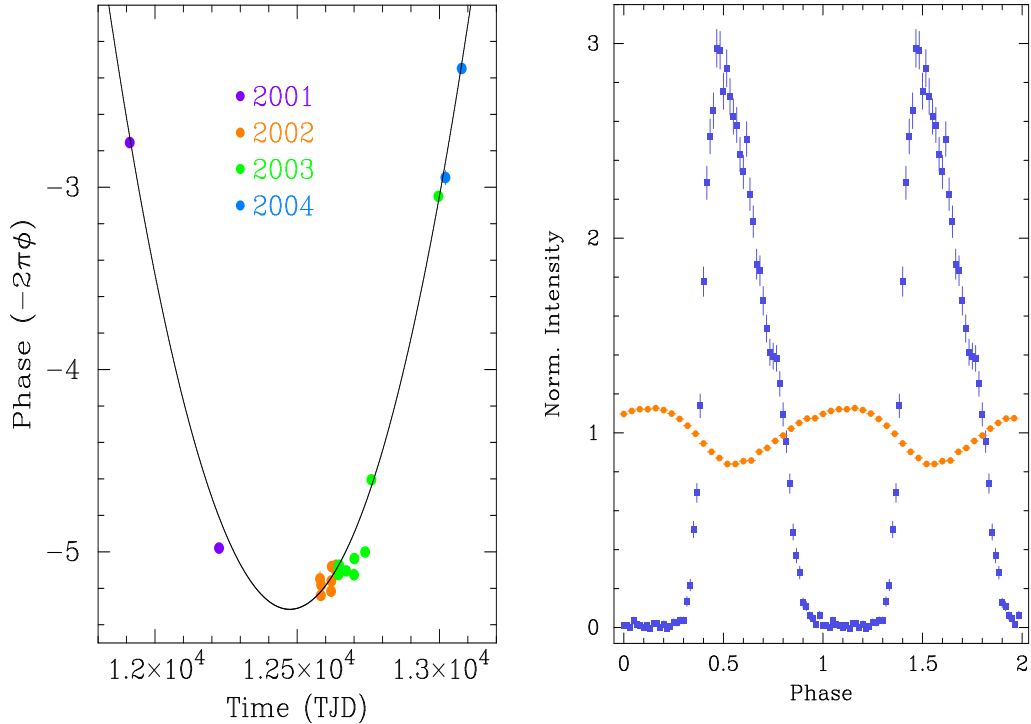

   \centering
   \resizebox{\hsize}{!}{\includegraphics{phase_noppdot.eps}
   \includegraphics{xo_jun04.eps}}
    \caption{Left panel: Results of the phase fitting technique used
to infer the P-\.P coherent solution for \src: the linear term (P
component) has been corrected, while the quadratic term (the \.P
component) has been kept for clarity.  The best \.P solution inferred
for the optical band is marked by the solid fit line.  Right panel:
1994--2002 phase coherently connected X--ray folded light curves
(largest pulsed fraction) of \src, together with the 2001-2004 phase
connected folded optical light curves. Two orbital cycles are reported
for clarity. }
        \label{timing}
    \end{figure*}
%------------------------------------------------------------------
The VLT spectral study revealed a blue continuum with no intrinsic
absorption lines \citepalias{io02b}. Broad ($\rm FWHM\sim 1500~\rm
km~s^{-1}$), low equivalent width ($EW\sim -2\div-6$ \AA) emission
lines from the He~II Pickering series (plus additional emission lines
likely associated with He~I, C~III, N~III, etc.) were instead
detected. These findings, together with the period stability and
absence of any additional modulation in the 1\,min--5\,hr period
range, are interpreted in terms of a double degenerate He-rich binary
\citep[a subset of the AM CVn class; see][for a review]{warn95} with
an orbital period of 321\,s, the shortest ever recorded. Moreover,
\src\ was noticed to have optical/X-ray properties similar to
those of \srcm, a 569\,s modulated soft X-ray source proposed
as a double degenerate system \citep{ram00,ram02b}.

More recently, \citet{hak03,hak04} reported on the detection of
spin--up, at a rate of $\sim$6.2$\times$10$^{-11}$\, s~s$^{-1}$, for
the 321\,s orbital modulation, based on optical data taken form the
Nordic Optical Telescope (NOT) and the VLT archive, and by using
incoherent timing techniques. Similar results were reported also by
\citet{stro03} for the X-ray data (ROSAT and Chandra) of \src\
spanning over 10 years and based on the \citet{hak03} results.
Finally, \citet{io03b} carried out a nearly-simultaneous Chandra/VLT
observational campaign reporting the presence of nearly
anti correlation between the optical and X--ray modulation. The X-ray
spectral study also allowed to characterise the emission mechanism; a
black body with a temperature of $\sim$60\,eV \citep{io03b}.

The study of \src\ and \srcm\ has posed, in the last years, serious
questions about their possible origin \citep[see][for a review of the
proposed theoretical models]{crop03}. Moreover, there is a number of
additional nice reasons for studying \src\ and the related
objects. Among other are the study of the gravitational wave, and the
possibility that \src\ is a progenitor of the He-accreting DDBs also
named AM CVns.  For this reason we requested and performed in the last
3.5 years several observations ranging from IR to X-ray band in order
to unveil the nature of \src\ and, correspondingly, of \srcm. Here we
report on the results obtained from (a) the VLT/TNG time-resolved
photometric monitoring aimed at accurately measure the P and \.P of
\src\ based on coherent timing techniques, (b) VLT-FORS1 polarimetry
observations, and (c) an \xmm\ pointing.

\section {Optical Observations}

After the successful first optical time-resolved observations of \src\
during 2001 (January 1st at TNG, and November 11th), we started a
relatively long-term project aimed at monitoring the source 321\,s
modulation. We obtained 21 pointings between November 2002 and May
2004 (11 at VLT and 10 at TNG) scheduled in a way such that it was
possible to keep the phase coherence among observations (the first
observations were obtained at 1st-2nd nights, 9-10th nights, 19-20th
nights, 49-50th nights, etc.). Four different optical bands (B, V, R
and I) have been used in order to study and monitor the pulse shape
and pulsed fraction as a function of wavelength and time (left panel
of figure~\ref{images} shows the R-band image of the region around 
\src).

Such a strategy resulted quite efficient in reaching the purposes of
the timing analysis, and allowed us also to extend the coherent
solution backward to the 2001 optical observations. The best optical
solution we found for P-\.P is for P=321.53040(4)\,s,
\.P=-3.6(1)$\times$10$^{-11}$\,s~s$^{-1}$ \citep[90\% uncertainties
are reported; for more details see][see also figure~\ref{timing}, left
panel]{io04}. Moreover, we found a slightly energy--dependent pulse
shape with the pulsed fraction increasing toward longer wavelengths,
from $\sim$12\% in the B-band to nearly 14\% in the I-band
\citep{io04}.

%______________________________________________ 
   \begin{figure}[bt]
   \centering
   \resizebox{\hsize}{!}{\rotatebox[]{-90}{\includegraphics{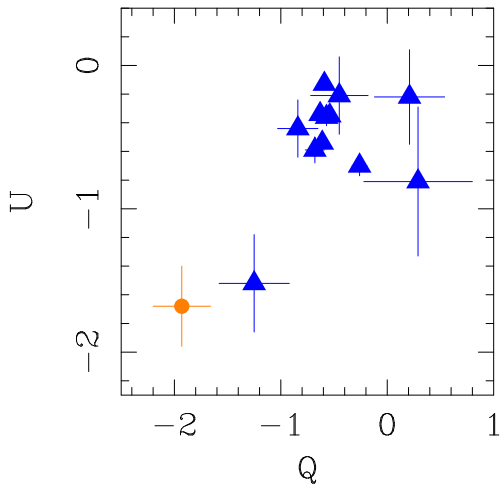}}}
%   %\includegraphics{empty.eps}
%   %\includegraphics{empty.eps}
   \caption{ Q and U Stokes parameters for all the 
objects detected within 3\arcmin\ from the \src\ position (on-axis).
%More details on the FORS1 spurious
%linear polarization effect can be found at
%www.eso.org/instruments/fors/pola.html.  
Values inferred for \src\ are clearly marked
(filled circle).}
              \label{QU}%
    \end{figure}
%______________________________________________________________
%______________________________________________________________
   \begin{figure}[th]
   \centering
   \resizebox{\hsize}{!}{\rotatebox[]{-90}{\includegraphics{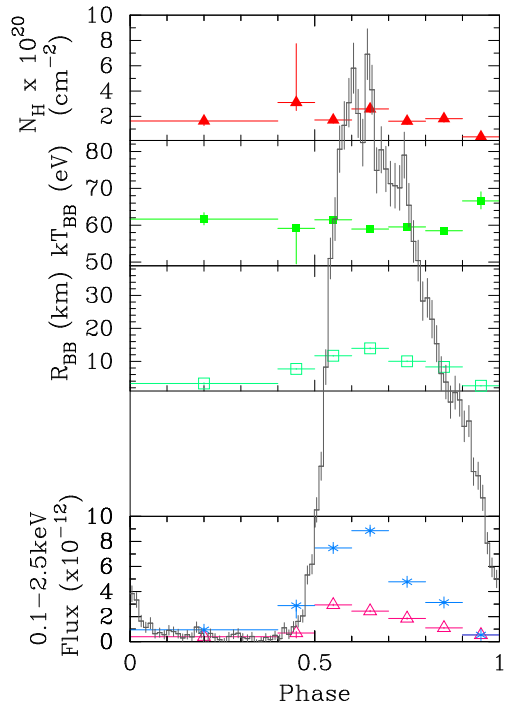}}}
%   %\includegraphics{empty.eps}
%   %\includegraphics{empty.eps}
   \caption{The results of the \xmm\ phase-resolved spectroscopy (PRS)
analysis for the absorbed blackbody spectral parameters: absorption,
blackbody temperature, blackbody radius (assuming a distance of
500\,pc), and absorbed (triangles) and unabsorbed (asterisks)
flux. Superposed is the folded X-ray light curve.}
              \label{xmm}%
    \end{figure}
%______________________________________________________________

The relatively high accuracy obtained for the optical phase coherent
P-\.P solution (in the January 2001 - May 2004 interval) allowed us to
extrapolate it backward to the ROSAT observations without loosing the
phase coherency, i.e.  only one possible period cycle consistent with
our P-\.P solution. Given the wider time interval spanned by the
X--ray observations (9.5 years) an even more accurate solution was
possible.  After taking into account the ROSAT photon arrival time
spacecraft clock time - UTC correction (in order to compare the ROSAT
data with the Chandra and \xmm\ observations), our best (optically
selected) X-ray solution is P=321.53033(2)\,s,
\.P=-3.67(1)$\times$10$^{-11}$\,s~s$^{-1}$ \citep[for more details
see][]{io04}. Figure~\ref{timing} (right panel) shows the optical
(2001-2004) and X--ray (1994-2002) light curves folded by using the
above reported P-\.P coherent solution, confirming the amazing
stability of the X--ray/optical anti correlation first noted by
\citet{io03b}.

In February 2003 we also collected VLT FORS1 polarimetry data for the
optical counterpart to \src\ in the B-band. Due to unforeseen large
FORS1 overheads only phase-resolved linear polarisation observations
were obtained.  In figure~\ref{QU} we reported the U and Q Stokes
parameters for all the detected objects within 3\arcmin\ from the
\src\ optical counterpart position (objects at larger off--axis angles
are indeed affected by the presence of spurious polarisation: more
details on this problem can be found at
www.eso.org/instruments/fors/pola.html). \src\ was found to be
linearly polarised at a level of $\sim$2.0$\pm$0.3\% (after correcting
the the field average polarisation, $\sim$0.7\%). It is worth noting
that the marginal detection of circular polarisation, at level of
about 0.5\%, has been recently reported by \citet{rei04}, and ascribed 
to a $\sim$10$^{6}$\,Gauss magnetic field.

\section{The \xmm\ observation}
During 2001 a Chandra observation of \src\ was obtained; in that case
the source was found to be more intense than expected resulting in a
high (and phase dependent) number of piled--up photons \citep[the
results of the Chandra observation are reported in][]{io03b}.  On 2002
November 1st we had the chance to observe \src\ with the \xmm\
instrumentations for about 26000\,s (see figure~\ref{images}, right
panel). This ensured us a dataset of unpiled--up photons allowing us
to increase the spectral analysis accuracy. Figure~\ref{xmm} shows the
results of the phase-resolved spectroscopic study we carried out
\citep[similar to the analysis reported in Figure~\ref{xmm} of][]{io03b}. With
respect to the latter analysis, the \xmm\ data show a lower value of
the absorption column, a relatively constant black body temperature, a
smaller black body size, and, correspondingly, a slightly lower
flux. All these differences may be ascribed to the pile--up effect in
the Chandra data, even though we can not completely rule out the
presence of real spectral variations as a function of time. In any case
we note that this result is in agreement with the idea of a
self-eclipsing (due only to a geometrical effect) small, hot and
X--ray emitting region on the primary star. Timing analysis did not
show any additional significant signal at periods longer or shorter
than 321.5\,s, (in the 5hr-200ms interval).  Results from the \xmm\
observation are more diffusely reported by \citet{io04}.

\section{Discussion}
In this contribution we briefly listed the main results obtained
thanks to a four years optical and X-ray monitoring of \src, a DDB
with an orbital period of only 5.4min. Even though the optical
monitoring is still active (in fact is quite important to continue 
the \.P study to look for variations or, more interestingly, a d\.P/dt 
component) a number of important implications can be considered.

It is now assessed that the orbital period is decaying  at a rate 
that is nearly consistent with that predicted by the ultracompact binary 
model in which the loss or orbital angular momentum is dominated by 
gravitational radiation and in which there is no mass exchange between 
the two stars \citep[see also][]{stro03}. In this respect the unipolar 
inductor model proposed by \citet{wu02} is a possible scenario to account 
for the observed X--ray flux from \src.

The X--ray and optical emissions are anti correlated at a level of
stability (no variations on a baseline of at least 4 years) which is
only observed in phase-locked binary systems. The anti correlation is
by far and so far the strongest (indirect) suggestion of the binary
nature of the 321\,s, where the X--rays illuminate the companion
surface originating the phase--shifted optical modulation.

The detection of linear and circular polarisation might imply the
presence of a relatively faint magnetic field, however we note that
the low polarisation level is also consistent with other possibilities, 
such as the Thomson scattering (sometimes observed in eclipsing binaries).

\begin{acknowledgements}
GianLuca Israel (GLI) is deeply grateful to Paul Plucinsky, Guenther
Hasinger, Piero Rosati, Roberto Gilmozzi and Martino Romaniello for
their help in planning and executing of the 2001 \AXAF/VLT
observational campaign. GLI also thanks Rosario Gonzalez--Riestra for
his help in setting the successful 2002 \xmm/VLT observational
campaign. GLI is also deeply indebted with Leonardo Vanzi and Stephen
Potter for their nice advices concerning the polarimetry
observations. GLI is also grateful to the VLT and TNG Team, in
particular to Marco Pedani (TNG), for their effort in optimising and
executing the FORS1 and DoLoRes observations of \src\ during the last
4 years !

This work is Based on observations made with the Italian Telescopio
Nazionale Galileo (TNG) operated on the island of La Palma by the
Centro Galileo Galilei of the INAF (Istituto Nazionale di Astrofisica)
at the Spanish Observatorio del Roque de los Muchachos of the
Instituto de Astrofisica de Canarias
This work was supported by COFIN grants from the Ministero
dell'Universit\`a e Ricerca Scientifica e Tecnologica (MURST), and by
the Italian Space Agency (ASI).

\end{acknowledgements}

\bibliographystyle{aa}

\begin{thebibliography}{}


%\bibitem[Bautz et al.(1998)]{bau98} Bautz, M. W. et al. 1998, SPIE, 3444, 210 
\bibitem [Beuermann et al. (1999)]{beu99} Beuermann, K., Thomas, H.-C., Reinsch, K., Schwope, A. D.,
Tr\"{u}mper, J, Voges, W., 1999, \aap, 347, 47
%\bibitem [Bowyer et al. (1991)]{bow91} Bowyer, S., \& Malina, R. F. 1991, in ``Extreme Ultraviolet Astronomy'', eds. R. F. Malina and S. Bowyer (New York: Pergamon Press), 397
\bibitem [Burwitz \& Reinsch (2001)]{bur01} Burwitz. V., Reinsch, K. 2001, {\it X-ray astronomy : stellar
endpoints, AGN, and the diffuse X-ray background}, Bologna, Italy, eds
White, N. E., Malaguti, G., Palumbo, G., AIP conference proceedings,
599, 522
%\bibitem [Campana et al (1998)]{camp98} Campana, S., Colpi, M., 
Mereghetti, S., Stella, L., \& Tavani, M.\ 1998, \aapr, 8, 279 
\bibitem [Cropper et al. (1998)]{crop98} Cropper, M., Harrop-Allin, M. K., Mason, K. O., Mittaz,
J. P. D., Potter, S. B., Ramsay, G., 1998, \mnras, 293, L57
\bibitem [Cropper (2003)]{crop03} Cropper, M., Ramsay, G., Wu, K. 2003, ASP Conf. Ser., Proceedings of Cape Town Workshop on magnetic CVs, December 2002, (see also astro-ph/0302240)
%\bibitem [Davies (2001)]{dav01 } Davis, J.E.\ 2001, \apj, 562, 575
%\bibitem [Dickey et al () (2001)]{dic01} Dickey, J.M.\& Lockman, F.J. 1990, \araa, 28, 215
%\bibitem[Garmire (1997)]{gar97} Garmire, G.~P.\ 1997, Bulletin 
of the American Astronomical Society, 29, 823  
%\bibitem [Haberl \& Motch (1995)]{hab95} Haberl, F., Motch, C. 1995, \aap, 297, L37
\bibitem [Hakala et al. (2003)]{hak03} Hakala, P., Ramsay, G., Wu, K., Hjalmarsdotter, L., J\"arvinen, S., 
J\"arvinen, A., Cropper, M. 2003, MNRAS, 343, L10
 \bibitem [Hakala et al. (2004)]{hak04} Hakala, P., Ramsay, G., Byckling, K. 2004, MNRAS, in press, (astro--ph/0405548)
\bibitem [Israel et al. (1999)]{io99} Israel, G.L., Panzera, M.R., Campana, S., Lazzati, D., Covino, S.,
Tagliaferri, G. 1999, \aap, 349, L1 (I99)
\bibitem [Israel et al. (2002a)]{io02a} Israel, G.L., Stella, L., Hummel, W., Covino, S., \&
Campana, S.\ 2002a, IAU Circ., 7835
\bibitem [Israel et al. (2002b)]{io02b} Israel, G.L. et al. 2002b, \aap, 386, L13 (I02)
\bibitem [Israel et al. (2003a)]{io03} Israel, G.L., Stella, L., Covino, S., Campana, S.,
Marconi, G., Mauche C.H., Mereghetti, S., Negueruela, I. 2003a, ASP
Conf. Ser., Proceedings of Cape Town Workshop on magnetic CVs,
December 2002, (see also astro-ph/0303124)
\bibitem [Israel et al. (2003b)]{io03b} Israel, G.L. et al. 2003b, \apj, 598, 492 
\bibitem [Israel et al. (2004)]{io04} Israel, G.L. et al. 2004, \aap, submitted 
%\bibitem [Marsh & Steeghs (2002)]{mar02} Marsh, T., Steeghs, D. 2002, \mnras, 331, L7
%\bibitem [Nelemans et al. (2001)]{nel01} Nelemans, G., Yungelson, L.R. \& Portegies Zwart, S.F. 2001, \aap, 
375, 890
%\bibitem [Norton et al. (2004)]{nor04} Norton, A. J., Haswell, C. A., Wynn, G. A. 2004, MNRAS, in press, (astro-ph/0206013)
%\bibitem [Phinney (2003)]{phin03} Phinney, E. S.\ 2003, AAS/High Energy Astrophysics Division, 35, 27.03  
\bibitem [Reinsch et al. (2004)]{rei04} Reinsch, K., Burwitz, V., and Schwarz R. 2004, {\it IAU Colloquium 194: Compact binaries in the Galaxy and Beyond}, RevMexAA, in press (astro--ph/0402458)
\bibitem [Ramsey et al. (2000)]{ram00} Ramsay, G., Cropper, M., Wu, K., Mason, K. O., Hakala, P. 2000,
\mnras, 311, 75
\bibitem [Ramsey et al. (2002a)]{ram02a} Ramsay, G., Hakala, P., Cropper, M. 2002a, \mnras, 332, L7
\bibitem [Ramsey et al. (2002b)]{ram02b} Ramsay, G., Wu, K., Cropper, M., Schmidt, G., Sekiguchi, K., Iwamuro, F., Maihara, T., 2002b, MNRAS, 333, 575
%\bibitem [Strohmayer (2002)]{stro02} Strohmayer, T., 2002, ApJ, 581, 577
\bibitem [Strohmayer (2003)]{stro03} Strohmayer, T., 2003, ApJ, 593, L39
%\bibitem [Strohmayer (2004)]{stro03} Strohmayer, T., 2004, ApJ, in press (astro--ph/0403675)
%\bibitem [Tavani \& London (1993)]{tav93} Tavani, M. \& London, R. 1993, \apj, 410, 281 
\bibitem [Warner (1995)]{warn95} Warner, B., 1995, Ap\&SS, 225, 249
%\bibitem [Warner (2002)]{warn02} Warner, B., Woudt, P. A. 2002, PASP, 792, 129
%\bibitem [Woudt (2003]{pat03} Woudt, P. A., Warner, B. 2003, in prep.
\bibitem [Wu et al. (2002)]{wu02} Wu, K., Cropper, M., Ramsay, G., Sekiguchi, K. 2002, \mnras, 331, 221


\end{thebibliography}

\end{document}